\documentclass{elsart5p}

\usepackage{graphics}
\usepackage{graphicx}
\usepackage{amssymb}

\newcommand{\beq}{\begin{equation}}
\newcommand{\eeq}{\end{equation}}
\newcommand{\barr}{\begin{eqnarray}}
\newcommand{\earr}{\end{eqnarray}}

\begin{document}

\begin{frontmatter}

\title{Critical behavior driven by the confining potential in optical lattices with ultra-cold fermions?}

\author[1,4]{V. L. Campo, Jr.}, \author[2]{J. Quintanilla}, \author[3]{C. Hooley}

\address[1]{Departamento de F\'{\i}sica, Universidade Federal de S\~ao Carlos, 13590-905, S\~ao Carlos, SP, Brazil}

\address[2]{ISIS Spallation Facility, STFC Rutherford Appleton Laboratory, Harwell Science and Innovation Campus, Didcot OX11 0QX, U.K.}

\address[3]{Scottish Universities Physics Alliance, School of Physics and Astronomy, University of St Andrews, \\ North Haugh, St Andrews, Fife KY16 9SS, U.K.}

\corauth[4]{Corresponding author. E-mail: vivaldo.leiria@gmail.com}

\begin{abstract}

A recent paper [V. L. Campo {\it et al.}, {\it Phys. Rev. Lett.} {\bf 99}, 240403 (2007)] has proposed a two-parameter scaling method to determine the phase diagram of the fermionic Hubbard model from optical
lattice experiments.  Motivated by this proposal, we investigate in more detail the behavior of the ground-state energy per site as a function of trap size ($L$) and confining potential ($
V(x) = t (x/L)^\alpha$) in the one-dimensional case. Using the BALDA-DFT 
method, we find signatures of critical behavior as $\alpha \to \infty$. 
\end{abstract}


\end{frontmatter}

\section{Introduction}

The Hubbard model was proposed in the 1960s as a simplified model of a system
of correlated fermions, capturing the competition between localisation due to strong interparticle
repulsion and itinerant behaviour due to intersite hopping.  It remains a model of central
importance in condensed matter physics; but despite this fact, and decades of investigative
effort, the model in dimensions $d>1$ remains unsolved, and its phase diagram is still a subject
of controversy.

Recently, a new line of attack has been proposed \cite{annphys_315_52}:\ to determine the phase diagram of the Hubbard
model not using theory, but using cold-atom experiments which --- unlike traditional condensed
matter systems --- should be essentially perfectly
described by the Hubbard Hamiltonian.  This is for two reasons.  Firstly, the lattice
in the cold-atom case is provided by a laser standing wave (or `optical lattice'), which can be made almost perfectly
periodic and sinusoidal.  Secondly, the interaction between the neutral atoms
should be well described by a contact interaction of the Hubbard form \cite{rmp_leggett,pet_smith}.  Other
advantages of such experiments include the following:\ the inter-atomic interaction strength
and hopping amplitude in these systems can be tuned over a wide range; the achievable particle numbers are
much greater than those available in current computer-based simulations of fermionic systems; and the total
energy (for example) is an easily measurable quantity.

However, a major difference between the solid-state and cold-atom realisations of the Hubbard model is the nature of
the particles' confinement.  In the solid-state case, it is of the `hard-wall' type, where the system is homogeneous except
for a very high potential step at its edges; in the cold-atom case, the confining potential arises from an external
electromagnetic field, and is almost always harmonic, leaving no residue of translational invariance.
How to extract the phase 
diagram of the homogeneous model from these inhomogeneous systems is 
discussed in Ref.~\cite{two_p_scaling}, where a two-parameter scaling method is 
proposed, with the two parameters $L$ and $\alpha$ representing respectively the size and shape of the confining potential.
The method involves considering different such shapes (i.e.\ different values of $\alpha$), and determining for each of them a thermodynamic ($L \to \infty$) limit
of some intensive quantity such as the energy per site.  Then one analyses the dependence of that quantity
on the shape of the confining potential and extrapolates (in $\alpha$) to the homogeneous (`hard-wall') limit.

It has been empirically observed \cite{two_p_scaling}, at least in the one-dimensional case, that the rate of convergence
to the thermodynamic limit is dependent on the shape of the confining potential.  Here, we address that question in more detail.  In the course of so doing, we find that the behavior during the second 
extrapolation, i.e.\ the extrapolation in the shape of the potential towards 
the homogeneous limit, displays signatures of critical behavior.

\begin{figure}
\begin{center}
\includegraphics[height=8.5cm, keepaspectratio, angle=-90]{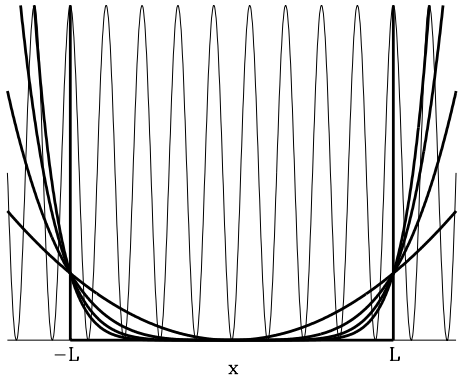}
\end{center}
\caption{Optical lattice and different confining potentials. The optical 
lattice with a high amplitude defines the sites and makes it possible to
describe the system by means of a Hubbard model, which is 
however inhomogeneous due to the confining potential (thick lines). As the exponent 
$\alpha$ in (\ref{vx_eq}) tends to $\infty$, the confining potential 
approaches the square well shape with width $2L$.} \label{fig1}
\end{figure}

\section{Model, and previously published results}

We consider the following inhomogeneous one-dimensional Hubbard model:
\barr
{\hat H} &=& -t\sum_{j,\sigma} ({\hat c}^\dagger_{j,\sigma} {\hat c}_{j+1,\sigma} + \mathrm{H.c.}) +
U\sum_j \hat{n}_{j,\uparrow} \hat{n}_{j,\downarrow} \nonumber \\
& & \quad + \sum_{j,\sigma} V(x_j)\hat{n}_{j,\sigma},  
\earr
where ${\hat c}^\dagger_{j,\sigma}$ creates a fermion with $z$-component of spin 
$\sigma$ at site $j$, $t$ measures the hopping integral between neighbouring 
sites, $U$ is the strength of the local interaction between two fermions of 
opposite spins, $\hat{n}_{j,\sigma} = {\hat c}^\dagger_{j,\sigma} 
{\hat c}_{j,\sigma}$ is the number operator for spin $\sigma$ at site $j$, and 
$V(x)$ is the external potential rendering the model inhomogeneous. The 
position of the site $j$ is $x_j = ja$, where $a$ is the lattice parameter of 
the chain, and $j$ takes any integer value. The external potential 
assumes the general form
\beq
V(x) = t\left|\frac{x}{L}\right|^\alpha \label{vx_eq}
\eeq
with $\alpha > 0$,
and therefore confines the fermions to a region of size 
$\sim 2L$, or more precisely, makes the probability of finding a fermion 
at $x$ with $|x|\gg L$ negligible. Accordingly, we refer to
$2L$ as the system size, and $2L/a$ as the number of sites, irrespective of the value of $\alpha$.
This allows us to define an intensive quantity as the corresponding extensive
quantity divided by $2L/a$.
In particular, we will be concerned
here with the energy per site, $Ea/2L$, and with the filling $f = Na/2L$, where 
$N$ is the number of fermions.   

The above Hubbard model, and its higher dimensional versions, can be realised
with trapped ultra-cold fermionic atoms in the presence of
an optical lattice with suitable laser intensity and 
wavelength (see Fig.~\ref{fig1}). The atoms are confined by an additional external 
potential whose exponent 
$\alpha$ is usually equal to $2$. In optical traps with Gaussian laser 
beams, a suitable superposition of two laser beams could eliminate the harmonic
part of the potential, leaving a quartic one. Such trap has already been reported \cite{x4}.
Similarly, superposing three or more laser beams, one could generate trap 
potentials with exponents of $6$, $8$, and so on. Despite the increasing 
experimental difficulties involved in making traps with higher exponents, the two-parameter
scaling method relies on this 
possibility. For applications to solid-state systems, one would be interested in the
thermodynamic limit ($L \to \infty$) of the above model with $\alpha = \infty$ 
(a square well potential of width $2L$).

\begin{figure}
\begin{center}
\includegraphics[height=8.5cm, keepaspectratio, angle=-90]{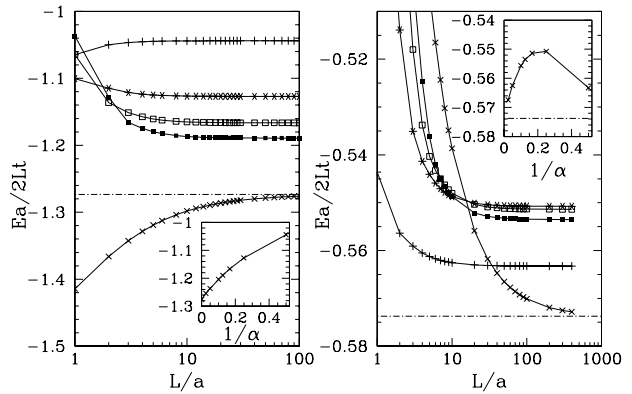}
\end{center}
\caption{Dependence of total energy on system size for different trap 
exponents, $\alpha = 2$ (${+}$), 4 ($\ast$), 6 ($\square$), 8 ($\blacksquare$)
 and $\infty$ (${\times}$). The filling is fixed at $f=1$. We show results for 
non-interacting fermions ($U=0$) on the left and for interacting fermions ($U=4t$) on the right.
Notice that for finite values of $\alpha$ the thermodynamic limit is attained with
 smaller system sizes than in the $\alpha=\infty$ case. Insets: the same quantity in
the thermodynamic limit plotted against $\alpha^{-1}$.  Figure reproduced from Ref.~\cite{two_p_scaling}.}\label{fig2}
\end{figure}

As 
Fig.~\ref{fig2} illustrates, given a finite exponent in the confining potential, 
simply 
increasing the size of the trap, $L$, is not enough to recover the homogeneous 
results. Traps with different exponents $\alpha$ have different thermodynamic 
limits of energy per site, and the same must happen with other properties.  The 
unavoidable confining potential with a finite exponent poses, therefore, an 
intrinsic obstacle to the extraction of the homogeneous model's phase 
diagram from such experiments.  To overcome it, one must make a sequence of experiments \cite{two_p_scaling}
with different exponents
and then extrapolate the results to get the limit
$\alpha \to \infty$.

\section{Method, and new results}

In order to achieve a better understanding of the role the trap exponent
$\alpha$ plays in the dependence of the energy per site on the system size, we
investigate both non-interacting ($U=0$) and interacting ($U \not = 0$) systems. While the 
former require only the diagonalization of a tridiagonal matrix,
the latter require methods which can deal with the strong
correlations coming from the interactions. We choose an approach based on 
Density-Functional Theory (DFT) \cite{dreizler_gross}, which has been specially 
developed to treat the one-dimensional inhomogeneous Hubbard model
\cite{neemias,balda_qmc}.  Its energy functional is 
constructed by means of the local density approximation based on the exact 
Bethe Ansatz solution for the homogeneous model:\ this approximation is termed
BALDA. Comparisons between BALDA and Quantum Monte Carlo (QMC) results
demonstrate that BALDA gives ground-state energies with an 
accuracy of a few percent \cite{balda_qmc}.

Our numerical simulations show that as the system size, $2L$, goes to infinity, the
system's ground-state energy per site approaches its thermodynamic limit 
according to
\beq
\epsilon(L) = \frac{a}{2L} E(L) = \epsilon_\infty + \left(\frac{2L}{\xi}\right)^{-\gamma}, \label{fss_eq}
\eeq
where $\epsilon_\infty$, $\xi$, and $\gamma$ all depend on the trap exponent $\alpha$ in a way which 
resembles critical behaviour, with the hard-wall case ($\alpha = \infty$) corresponding to the critical point.

\begin{figure}
\begin{center}
\includegraphics[height=8.5cm, keepaspectratio, angle=-90]{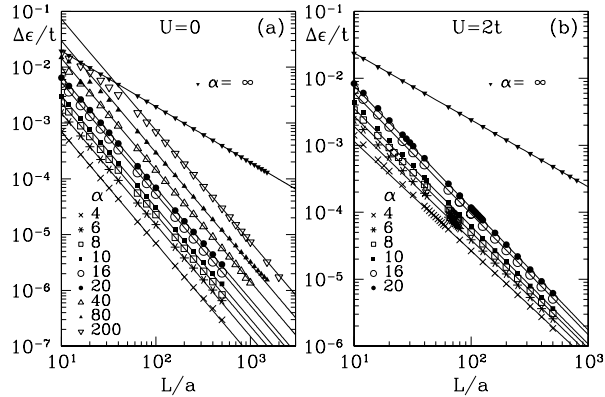}
\end{center}
\caption{Dependence of $\Delta \epsilon = E/2L - \epsilon_\infty$ on the
system size for (a) non-interacting fermions ($U=0$) and (b) interacting fermions ($U=2t$). In both cases, numerical data from systems with filling $f=0.5$ were used. For each value of the trap exponent, $\alpha$, Eq.~(\ref{fss_eq}) fits the data very well even for quite small systems. Similar behavior is found for fillings $f=1.0$ and $f=1.5$ and also interaction strengths $U=4t$ and $U=8t$.}\label{fig3}
\end{figure}

\begin{figure}
\begin{center}
\includegraphics[height=8.5cm, keepaspectratio, angle=-90]{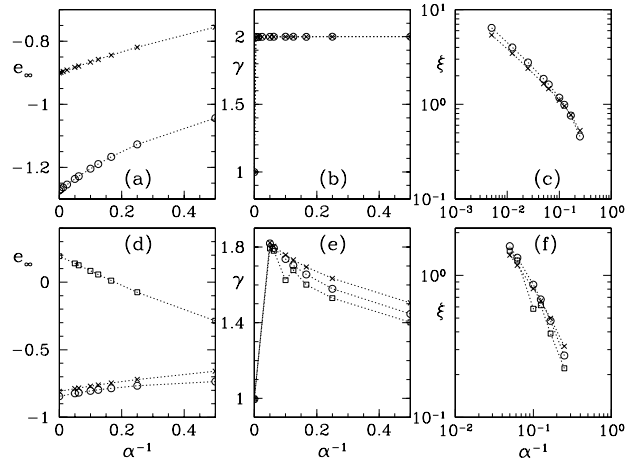}
\end{center}
\caption{Dependence of the thermodynamic limit of energy per site, $e_\infty$,
of the finite-size scaling exponent, $\gamma$, and of the correlation length, $\xi$, on 
the trap exponent, $\alpha$, for non-interacting ((a), (b), and (c)) and interacting 
((d), (e), and (f)) fermions. In the non-interacting case, we show results for 
fillings $f=0.5$ ($\times$) and $f=1.0$ ($\circ$); in the interacting 
($U=2t$) case, we show in addition results for $f=1.5$ ($\square$).
}\label{fig4}
\end{figure}

In Fig.~\ref{fig3}, we show that the energy per site follows 
Eq.~(\ref{fss_eq}) for both non-interacting and interacting systems. 
Through the fitting we can extract the dependence of its three 
parameters on the trap exponent (Fig.~\ref{fig4}). Although Fig.~\ref{fig3} only displays results for filling $f=0.5$, we have found similar 
behavior for different fillings ($f=1$ and $f=1.5$) and different 
interaction strengths ($U=4t$ and $U=8t$). 

Fig.~\ref{fig4} displays the dependence of the thermodynamic limit of 
energy per site $\epsilon_\infty$, the finite-size scaling exponent $\gamma$, and
the length $\xi$ on the trap exponent $\alpha$.
These graphs are reminiscent of critical behaviour, with $\alpha^{-1} = 0$ (i.e.\ the square-well trap) as the critical point.
While the 
thermodynamic-limit energy $\epsilon_\infty$ is a continuous function of 
$\alpha^{-1}$, the exponent $\gamma$ seems to be discontinuous at 
$\alpha^{-1} = 0$, and the length $\xi$ seems to diverge as
$\alpha^{-1} \to 0$ --- lending support to our use of the term `correlation length'.

Note that, in the interacting case, the finite-size scaling exponent $\gamma$
is $\alpha$-dependent, and furthermore seems to tend to the non-interacting value $\gamma=2$ as $\alpha^{-1} \to 0$.
Also, in all finite-$\alpha$ cases we have $\gamma > 1$.
This accords with our earlier observation that the thermodynamic 
limit is more easily achieved for finite trap exponents.
The reason for this is likely to be that for finite $\alpha$ the wave functions can
extend beyond $|x|=L$.  Thus the system does not end abruptly, which in turn avoids strongly
perturbing the fermionic fluid.

The apparent discontinuity in $\gamma(\alpha)$ at $\alpha^{-1} = 0$, however, remains surprising.
In the non-interacting case, we have 
performed calculations for $\alpha$ as large as $200$. Looking carefully at 
Fig.~\ref{fig3}a, one can see that in the case $\alpha = 200$, the 
points corresponding to smaller systems seem to follow a line whose 
$\gamma$ is equal to 1; but as the system size becomes larger, the 
energy starts to follow the line with $\gamma = 2$.  We can, therefore, 
identify a transition length, $L_T$, and we expect $L_T$ to become larger and 
larger as the trap exponent is increased, diverging when $\alpha^{-1} \to 0$. 
This implies a relation between $L_T$ and the correlation 
length $\xi$, though additional computations with larger systems 
are necessary to determine its nature.

The dependence on filling fraction, $f$, is also instructive. 
The finite-size scaling exponent $\gamma$ and the correlation length $\xi$ do depend on $f$;
however, the plots indicate that this dependence disappears as we approach the critical
point $\alpha^{-1} = 0$. 
This is exactly what we would expect in a critical scenario:\ the critical point
would be described only by its universality class, which would be independent of a 
parameter like the filling fraction. In particular, the slopes in 
Figs.~\ref{fig4}c and \ref{fig4}f are essentially the same for the different 
fillings and give us directly the critical exponent $\nu$ associated to the 
correlation length. We obtain $\nu \sim 0.51$ for non-interacting fermions 
and $\nu \sim 0.70$ for interacting fermions.  We must emphasise, however, that this latter is a rough
estimate only.  Computations with larger systems will be necessary to determine 
the true behavior of interacting fermions near the critical point, even if the BALDA method
is reliable in such extreme cases.
  
\section{Conclusions}

In conclusion, we have considered the inhomogeneous one-dimensional Hubbard model,
studying the evolution of the ground-state energy per site towards its
thermodynamic limit as the system size is increased. Looking at the role played
by the trap exponent, we have found an apparently critical behavior, which still
needs more detailed characterization.  While the results for non-interacting 
fermions are exact, the results for interacting fermions were obtained within 
the BALDA approximation.

Since the observed critical behavior has a geometrical 
origin, we expect that such behavior must happen independently of the interaction
 strength. Our treatment using BALDA can be interpreted as a mean field 
calculation, which is able to capture the critical behavior but gives inaccurate 
results for critical exponents, for example. To overcome this methodological 
limitation, one should adopt more accurate computational methods, such
as QMC \cite{rigol} or Density Matrix Renormalization Group (DMRG) \cite{rmp_dmrg}.  These
are, however, much more computationally intensive than a DFT calculation.

From the analytical point of view, even the non-interacting case deserves 
further exploration.  In particular, in the absence of interactions it should be easier to find a 
renormalization group approach with a convenient decimation scheme to extract 
the critical behavior accurately. Once such scheme has been found, given the 
similarity between the interacting and non-interacting results, such an approach 
could be tentatively generalized to the interacting case.

\section{Acknowledgments}

The authors would like to thank J. L. Cardy for having brought to our attention
the importance of a scaling law, and K. Capelle and J. M. F. Gunn for helpful 
discussions. CH acknowledges financial support from the EPSRC (UK), JQ 
acknowledges support by CCLRC (now STFC) in association with St. Catherine¡Çs 
College, Oxford and VC acknowledges support from Universidade Federal de Sao 
Carlos.

\end{document}